\documentclass[aps,preprint,a4paper,12pt,longbibliography,nofootinbib]{revtex4-1}

\usepackage{eurosym}
\usepackage{graphicx}
\usepackage{amsmath}
\usepackage{mathrsfs}
\usepackage{bm}
\usepackage{color}
\usepackage{amssymb}

\begin{document}

\title{A new mass scale, implications on black hole evaporation and holography}
\author{Piyabut Burikham}
\email{piyabut@gmail.com}
\affiliation{High Energy Physics Theory Group, Department of Physics,
Faculty of Science, Chulalongkorn University, Phyathai Rd., Bangkok 10330, Thailand}
\author{Rujikorn Dhanawittayapol}
\email{rujikorn.physics@gmail.com}
\affiliation{High Energy Physics Theory Group, Department of Physics,
Faculty of Science, Chulalongkorn University, Phyathai Rd., Bangkok 10330, Thailand}
\author{Taum Wuthicharn}
\email{Taum.W@student.chula.ac.th}
\affiliation{High Energy Physics Theory Group, Department of Physics,
Faculty of Science, Chulalongkorn University, Phyathai Rd., Bangkok 10330, Thailand}

\date{\today }

\begin{abstract}

We consider a new mass scale $M_{T}=(\hbar^{2}\sqrt{\Lambda}/G)^{1/3}$ constructed from dimensional analysis by using $G$, $\hbar$ and $\Lambda$ and discuss its physical interpretation.  Based on the Generalized Uncertainty Relation, a black hole with age comparable to the universe would stop radiating when the mass reaches a new mass scale $M'_{T}=c(\hbar/G^{2}\sqrt{\Lambda})^{1/3}$ at which its temperature corresponds to the mass $M_{T}$.  Black hole remnants could have masses ranging from a Planck mass to a trillion kilograms.  Holography persists even when the uncertainty relation is modified to the Minimum Length Uncertainty Relation~(MLUR).  The remnant black hole entropy is proportional to the surface area of the black hole in unit of the Planck area in arbitrary noncompact dimensions.  

\vspace{5mm}

{Keywords: mass scale, generalized uncertainty principle, black hole remnant, Hawking radiation, holography}

\end{abstract}

\maketitle

\section{Introduction}

Based on dimensional analysis, it is known that we can construct the Planck mass
\begin{equation}
\sqrt{\frac{\hbar c}{G}},
\end{equation}
from three fundamental constants, the gravitational constant, the Planck constant and the speed of light: $G$, $h$ and $c$ respectively.  The combination is determined uniquely and a number of physical interpretations are possible.  

By adding cosmological constant ($\Lambda$), we may construct two additional mass scales containing $\Lambda$, namely $M_W$ and $M'_{W}$ given by
\begin{equation}
\frac{\hbar}{c}\sqrt{\frac{\Lambda}{3}} \quad \text{  and  } \quad \frac{c^{2}}{G}\sqrt{\frac{3}{\Lambda}},
\end{equation}
respectively.  The two mass scales have been proposed by Wesson~\cite{Wesson:2003qn} and certain physical interpretations are discussed.  The first and smaller Wesson mass, $M_{W}$, can be thought of as the minimum mass scale in nature~\cite{Wesson:2003qn,Boehmer:2005sm}, whilst the second and large Wesson mass, $M'_{W}$, is interpreted as the mass of the visible universe~\cite{Wesson:2003qn,Burikham:2015nma}.  Another interesting interpretation of the Wesson masses in relation to the Hawking temperature can be found in Appendix \ref{app}.  We should observe that the first Wesson mass has no $G$ and the second Wesson mass has no $\hbar$.  

It can be shown that these mass scales are all connected by a dimensionless quantity
\begin{equation}
\left(\frac{c^{3}}{\hbar G \Lambda}\right)^{\frac{1}{2}}\equiv N^{1/2},
\end{equation}
which contains all four fundamental constants.  In Ref.~\cite{Burikham:2015nma}, it is argued that new mass scales $M_{\Lambda}\simeq \sqrt{M_{W}M_{P}}$ and $M'_{\Lambda}\simeq \sqrt{M_{P}M'_{W}}$ sitting logarithmically in the middle between the two pairs of masses, $(M_{W}, M_{P})$ and $(M_{P}, M'_{W})$ can be interpreted as the minimum masses of {\it static} classical object having classical radius larger than its own Compton wavelength.  Interestingly, the fundamental dimensionless quantity $N\sim 10^{120}$ can be interpreted to be the number of quantum gravity bits on the entire cosmic horizon area, i.e. $N\sim (\Lambda R_{p}^{2})^{-1}$.  This is the holographic bound on the maximum numbers of degrees of freedom we can put in the observable universe.   

A careful reader would notice that each of the ``fundamental'' masses, $M_{W}, M_{P}, M'_{W}$, is constructed from three out of four natural constants.  It means that we miss one other possible mass scale; the one without $c$.  By simply multiply $M_{W}$ with dimensionless factor $N^{1/3}$, we obtain up to a numerical factor
\begin{equation}
M_{T}=\left(\frac{\hbar^{2}\sqrt{\Lambda}}{G}\right)^{1/3},  \label{MT}
\end{equation}
remarkably a mass scale with no $c$\footnote{Actually in Ref.~\cite{Funkhouser:2007bp,Capozziello:2008zm} with some typos, the same mass was proposed but the $c$ independence is not apparent since the authors define $\Lambda$ to be scaled by $c$ in the unit of time.  The physical interpretations are also different from ours here.}.  Interestingly, we might divide $M'_{W}$ by $N^{1/3}$ to obtain another new mass up to a numerical factor
\begin{equation}
M'_{T} = c\left( \frac{\hbar}{G^{2}\sqrt{\Lambda}}\right)^{\frac{1}{3}}. \label{MT'}
\end{equation}
This mass is introduced, upto a dimensionless factor with different motivation, in Ref.~\cite{Burikham:2015sro}.

In SI unit, the value of $M_{T}$ and $M'_{T}$ are approximately $10^{-28}$ and $10^{12}$ kg respectively.  Note that $M_{T}$ is roughly one-tenth of the proton mass.

These mass scales may be plotted into a logarithmic scale as follows,
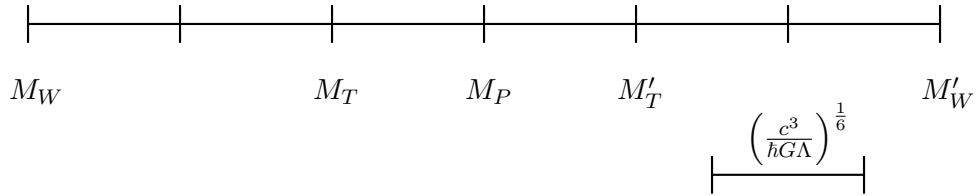
\begin{figure}[h]
\centering
\setlength{\unitlength}{1cm}
\thicklines
\begin{picture}(20,5)
\put(2.5,3){\line(1,0){12}}
\put(2.5,2.75){\line(0,1){0.5}}
\put(2.25,2){$M_{W}$}
\put(4.5,2.75){\line(0,1){0.5}}
\put(6.5,2.75){\line(0,1){0.5}}
\put(6.25,2){$M_{T}$}
\put(8.5,2.75){\line(0,1){0.5}}
\put(8.25,2){$M_{P}$}
\put(10.5,2.75){\line(0,1){0.5}}
\put(10.25,2){$M'_{T}$}
\put(12.5,2.75){\line(0,1){0.5}}
\put(14.5,2.75){\line(0,1){0.5}}
\put(14.25,2){$M'_{W}$}
\put(11.5,1){\line(1,0){2}}
\put(11.5,0.75){\line(0,1){0.5}}
\put(13.5,0.75){\line(0,1){0.5}}
\put(11.95,1.375){$\left(\frac{c^{3}}{\hbar G\Lambda}\right)^{\frac{1}{6}}$}
\end{picture}
\caption{Hierachy of masses on the logarithmic scale.}
\end{figure}\label{fig1}

From Fig.~\ref{fig1}, the geometric relation between mass scales are apparent.  The Planck mass is given by the geometric mean $M_{P}=\sqrt{M_{W}M'_{W}}$ of the two Wesson masses.  The new mass scale without $c$ is given by another geometric mean $M_{T} \simeq (M_{W}M_{P}^{2})^{1/3}=(M_{W}^{2}M'_{W})^{1/3}$, dividing each pair of the mass scales $(M_{W}, M_{P})$ and $(M_{W}, M'_{W})$ into one-third the values on the log scale. 

The motivation of this paper is to find physical meaning for the mass scales $M_{T}, M'_{T}$.  In the next Section, we will show that a form of generalized uncertainty relation could lead to a physical interpretation of the new mass scales $M_{T}, M'_{T}$.  We also discuss the consequences of the uncertainty relation to the Hawking radiation.

\section{Generalized uncertainty principle~(GUP)}  \label{SectGUP}

In this section, we will derive the General Uncertainty Relation in some particular forms and discuss the connection to the mass scales $M_{T},M'_{T}$.  There is a number of good review papers on the GUP~(\cite{Garay:1994en,Kempf:1994su,Tawfik:2015rva,Hossenfelder:2012jw}) where the reader can consult.  A number of different interpretations and derivations of the GUP is also available in those references.    The consequences to the physics of black hole radiation will be explored in subsequent section.

We are all familiar with Heisenberg's uncertainty principle,
\begin{equation}
	\Delta x \Delta p \ge \frac{\hbar}{2}.  \nonumber
\end{equation}
The relation gives a natural limit on how precise we can measure position and momentum of a quantum particle even in the absence of a real measurement.  From this relation, it is possible to have zero uncertainty in position ($\Delta x =0$) when we have infinite uncertainty in momentum.  The uncertainty relation in this original form reflects the wave nature of quantum particle at an instantaneous moment in time.  

Generally, however, there will be other factors which increase uncertainty of the measurement.  If the object or particle under measurement interacts with its own energy/momentum, there will be additional uncertainty proportional to $\Delta p~(\Delta x \ge \beta\Delta p)$ at the leading order.  An example is gravitation where the particle's own gravity gives the minimum position uncertainty of the same size as its Schwarzschild radius.  Another uncertainty originates from the size of the probe~($\Delta x \ge \ell$) we use in the measurement.  Thus,
\begin{equation}
	\Delta x \Delta p \ge \frac{\hbar}{2}+\ell\Delta p+\beta(\Delta p)^{2}\nonumber
\end{equation}
gives a generalized uncertainty relation.  Previously, $\Delta x$ can have a value as small as zero, but with the additional uncertainties it is clear that when $\Delta p \to \infty$, $\Delta x$ also diverges to infinity.  Hence, there is a non-zero minimum value of position uncertainty $\Delta x_{min}$ which can be calculated from the generalized uncertainty relation.  Consider
\begin{eqnarray}
	\Delta x &\ge& \frac{\hbar}{2\Delta p}+\ell+\beta\Delta p\nonumber.
\end{eqnarray}
By minimizing $\Delta x$ with respect to $\Delta p$, we obtain the uncertainty in momentum that minimizes $\Delta x$,
\begin{eqnarray}
	\Delta p_{c} &=& \sqrt{\frac{\hbar}{2\beta}},  \nonumber  \\
	\therefore \Delta x_{min} &=& \ell+\sqrt{2\hbar\beta}. \nonumber
\end{eqnarray}
Certainly we can take the probe~(particle) size to be extremely small, e.g. by using short wavelength photon for the measurement.  Therefore the $\ell$ term can be taken to be almost arbitrarily small~(limit on $\ell$ itself is naturally $\sqrt{2\hbar \beta}$) and thus $\Delta x_{min}$ is given roughly by the second term $\sqrt{2\hbar \beta}$.  

One generalization of the uncertainty relation which can be expressed in the above form is called the `Minimum Length Uncertainty Relation'~(MLUR),
\begin{equation}
	\Delta x\ge\frac{\hbar}{2\Delta p}+\frac{2R\Delta p}{Mc}   \label{MLUReqn}
\end{equation}
The additional term comes from e.g. the following thought experiment~\cite{Salecker:1957be,Hossenfelder:2012jw}; suppose we have a particle traveling parallel to a mirror. Let $R$ be the distance between the particle and the mirror. The time it would require for photon to be emitted from the particle, reflected by the mirror and absorbed back by the same particle is roughly $t=2R/c$. During this time, the particle would have acquired an additional position uncertainty $\Delta x=(\Delta p/M)t=2R\Delta p/Mc$.  Note that the MLUR gives the string-inspired GUP~\cite{Veneziano:1986zf,Gross:1987ar,Amati:1988tn,Konishi:1989wk,Adler:1999bu,Adler:2001vs} when $R\to R_{p}, M\to M_{p}$.

If we minimize $\Delta x$ with respect to $\Delta p$ from the MLUR (\ref{MLUReqn}), we obtain
\begin{equation}
\Delta x_{min} = 2\sqrt{\frac{R\hbar}{Mc}}.   \label{dxmin}
\end{equation}

An alternative derivation of the minimum uncertainty (\ref{dxmin})~(up to some numerical factor) is to consider the measurement of the position of the same particle twice in the Heisenberg picture, this would yield uncertainty in the distance travelled by that particular particle,
\begin{equation}
	\hat{x}(t)=\hat{x}(0)+t \frac{d\hat{x}}{dt}.\nonumber
\end{equation}
From $d\hat{x}(t)/dt=i[\hat{H},\hat{x}(t)]=\hat{p}/M$, we get
\begin{equation}
	\hat{x}(t)=\hat{x}(0)+\hat{p}(0)\frac{t}{M}.\nonumber
\end{equation}
With this relation, we find that,
\begin{eqnarray}
	\because \Delta A \Delta B&\ge&\frac{1}{2i}\langle[\hat{A},\hat{B}]\rangle, \nonumber \\
	\left[\hat{x}(0),\hat{x}(t)\right]&=&i\hbar\frac{t}{M}, \nonumber \\
	\Delta x(0)\Delta x(t)&\ge&\frac{t\hbar}{2M}.\nonumber
\end{eqnarray}
If $R$ is an apparatus size which we use to measure position of the particle, the observant time is related to $R$ by $t\sim 2R/c$~(if we use photon to measure the position),
\begin{eqnarray}
	\therefore\Delta x(0)\Delta x(t)&\ge&\frac{R\hbar}{Mc},\nonumber \\
	\Delta x&\ge&\sqrt{\frac{R\hbar}{Mc}},\nonumber
\end{eqnarray}
the same result as Eqn.~(\ref{dxmin}) up to a numerical factor.  The physical meaning of the above uncertainty is the following.  Position uncertainty of a particle increases as time elapses.  The time dependence of the uncertainty is $\sqrt{t}$, distinctive characteristic of the random walk.  It should be emphasized that this is the total uncertainty from intrinsic uncertainty due to the wave nature of the particle and the uncertainty from the time elapse required by the measurement process of the apparatus with size $R$.     

Next we incorporate uncertainty from gravity into the uncertainty relation.  Since our particle can not be smaller than its Schwarzschild radius,
\begin{equation}
	\Delta x \ge R_{S} = \frac{2GM}{c^{2}}.\nonumber
\end{equation}
By adding both kinds of position uncertainty together, we get,
\begin{equation}
	\Delta x\ge\sqrt{\frac{R\hbar}{2Mc}}+\frac{GM}{c^{2}}\nonumber
\end{equation}
From this MLUR, we find
\begin{eqnarray}
\Delta x_{min}&=&\frac{3}{2}\left(\frac{RG \hbar}{c^{3}}\right)^{\frac{1}{3}}=\frac{3}{2}(R R_{p}^{2})^{1/3},	\\
M_{c}&=&\frac{1}{2}(R\sqrt{\Lambda})^{1/3}M'_{T}=\frac{1}{2}\left(\frac{R}{R_{W}}\right)^{1/3}M'_{T},   \label{Mmin}
\end{eqnarray}
where $M_{c}$ is the corresponding mass which gives $\Delta x_{min}$, $R_{W}\equiv \sqrt{1/\Lambda}$ and $R_{p}=\sqrt{\hbar G/c^{3}}$ is the Planck length.

We may rearrange $\Delta x_{min}$ into another form,
\begin{equation}
	\Delta x_{min} = \frac{3}{2}(R R_{p}^2)^{1/3}=\frac{3M_{T}}{2M'_{T}}\left(\frac{R}{\Lambda}\right)^{\frac{1}{3}}\nonumber
\end{equation}
We can define the corresponding Compton radius for each mass scale as $R_{i}=\hbar/M_{i}c$.  Since $R_{W}=\Lambda^{-1/2}$ and $R_{W'}=G\hbar\sqrt{\Lambda}/c^{3}\sqrt{3}$, other possible forms of $\Delta x_{min}$ are
\begin{equation}
	\Delta x_{min}=\frac{3M_{T}}{2M'_{T}}(R R_{W}^{2})^{1/3}\simeq (R R_{W}R_{W'})^{1/3}=\left(\frac{R}{R_{W}}\right)^{1/3}R_{T},  
\end{equation}
where $R_{T}$ is the Compton radius of the new mass scale $M_{T}$.  If we set $R$ equal to the size of observable universe $R_{W}$, then
\begin{equation}
	\Delta x_{min}=R_{T}=(R_{W}R_{p}^{2})^{1/3}.
\end{equation}
This means $R_{T}$ is the minimum uncertainty of distance travel by a particle over the size of the observable universe.  Alternatively, we can interpret $R_{T}$ to be the minimum position uncertainty of the particle when the apparatus size is radius of the observable universe $R_{W}$.  According to (\ref{Mmin}), the corresponding mass that would yield this $\Delta x_{min}$ is $M'_{T}$.  Therefore, a possible physical interpretation of $M_{T}$ is the mass whose the Compton radius $R_{T}=(R_{p}^{2}R_{W})^{1/3}$ is the minimal uncertainty of a particle travel over the entire de Sitter radius $R_{W}$.  In other words, the {\it maximally possible} minimum uncertainty of position of a particle in the universe is the Compton radius of the mass $M_{T}$.  On the other hand, the {\it minimally} possible minimum uncertainty $\Delta x_{min}$ is when we set $R=R_{p}$ giving the absolute minimal $\Delta x_{min}=R_{p}$, the Planck length.  Numerically, $\Delta x_{min}$ from the MLUR is in the range of value from the Planck length $10^{-35}$ m to a few femtometers.    

Other interpretation for $R_{T}$ is the size of our universe if its density is the Planck density($\rho_{p}$),
\begin{eqnarray}
	\rho_{p}&=&\frac{M_{p}}{\frac{4}{3}\pi R_{p}^{3}}=\frac{M'_{W}}{\frac{4}{3}\pi r^{3}}\nonumber \\
		\therefore r&=&\left(\frac{M'_{W} R_{p}^{3}}{M_{p}}\right)^{1/3}=(R_{p}^{2}R_{W})^{1/3}=R_{T},\nonumber
\end{eqnarray}
where we assumed that the mass of the universe is $M'_{W}$.  If we compress the universe until it reaches the Planck density, its size will be in the order of $R_{T}$.  

\section{Minimum Length Uncertainty Relation and Black Hole Remnants}  \label{SectTH}

Generally, we believe that a black hole whose temperature is hotter than its surrounding would keep emitting black body radiation until there is nothing left.  This assumption is based on the viewpoint of miniature black hole as unstable quantum system.  However, if we apply the MLUR to black hole evaporation mechanism, it is possible to stop the process before the black hole can evaporate completely.  The existence of black hole remnants could lead to a possible resolution of the information loss paradox~(see Ref.~\cite{Chen:2014bva,Chen:2014jwq} and references therein).  

As a rough estimation, we can apply the Heisenberg's uncertainty principle to the black hole evaporation mechanism~\cite{Adler:2001vs}, starting from uncertainty relation,
\begin{equation}
	\Delta p\ge\frac{\hbar}{\Delta x}.\nonumber
\end{equation}
By setting $\Delta x$ equals to the Schwarzschild diameter,
\begin{equation}
	\Delta p=\frac{\hbar c^{2}}{4GM} \nonumber
\end{equation}
gives momentum uncertainty of particle radiated from a black hole.  The corresponding energy uncertainty is then $c\Delta p=\hbar c^{3}/4GM$.  If we identify this value as energy of the emitted photon, the corresponding temperature would agree with the Hawking temperature up to a factor of $2\pi$,
\begin{eqnarray}
	\because E&\simeq&k_{B}T,\nonumber \\
	\therefore T&\simeq&\frac{\hbar c^{3}}{4GMk_{B}}.\nonumber
\end{eqnarray}
Although this result agrees with Bekenstein-Hawking radiation, it does not prevent black hole from total evaporation. However, if we were to use the string-inspired Generalized Uncertainty Principle, the evaporation process would stop around the Planck mass~\cite{Adler:2001vs}. 

If instead of the conventional uncertainty relation, we adopt the MLUR relation (\ref{MLUReqn}).  We can use the new relation to calculate black hole temperature. We start by multiplying both sides of the equation with $\Delta p$ and treat it as a quadratic equation,
\begin{eqnarray}
	0&=&\frac{2R}{Mc}(\Delta p)^{2}-\Delta x\Delta p+\frac{\hbar}{2}\nonumber \\
	\therefore \Delta p&=&\frac{Mc\Delta x}{4R}\left(1\pm\sqrt{1-\frac{4R\hbar}{Mc(\Delta x)^{2}}}\right). \nonumber
\end{eqnarray}
Setting $\Delta x=4GM/c^{2}$, the momentum uncertainty is then
\begin{equation}
	\Delta p=\frac{GM^{2}}{Rc}\left(1\pm\sqrt{1-\frac{R\hbar c^{3}}{4G^{2}M^{3}}}\right). \nonumber
\end{equation}
By taking the large mass limit, if the negative sign is chosen, this result would agree with Hawking temperature with a calibrating factor of $\pi$.  The positive sign corresponds to no physical meaning.  Thus, the temperature of black hole is,
\begin{equation}
	T=\frac{GM^{2}}{\pi Rk_{B}}\left(1-\sqrt{1-\frac{R\hbar c^{3}}{4G^{2}M^{3}}}\right).  \label{TMeqn}
\end{equation}
By plotting temperature against mass, we can see that the temperature will continue to raise as the mass decreases, just as it should, until the value of $M$ rendering term under the square root sign negative and meaningless.  This is the point where black hole evaporation process would stop.
\begin{figure}[tbh]
\center
{\includegraphics[width=0.45\textwidth]{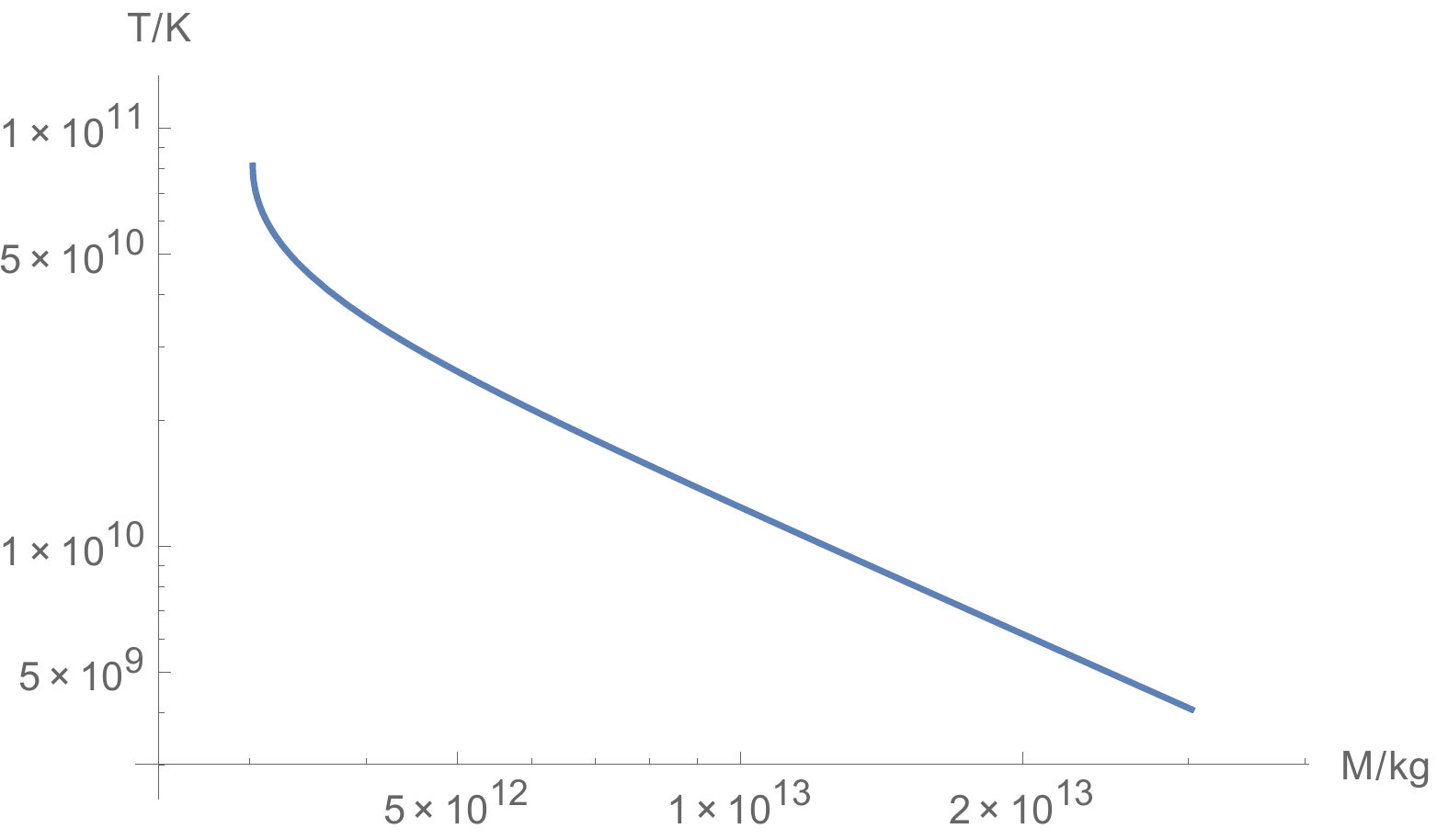}
\includegraphics[width=0.45\textwidth]{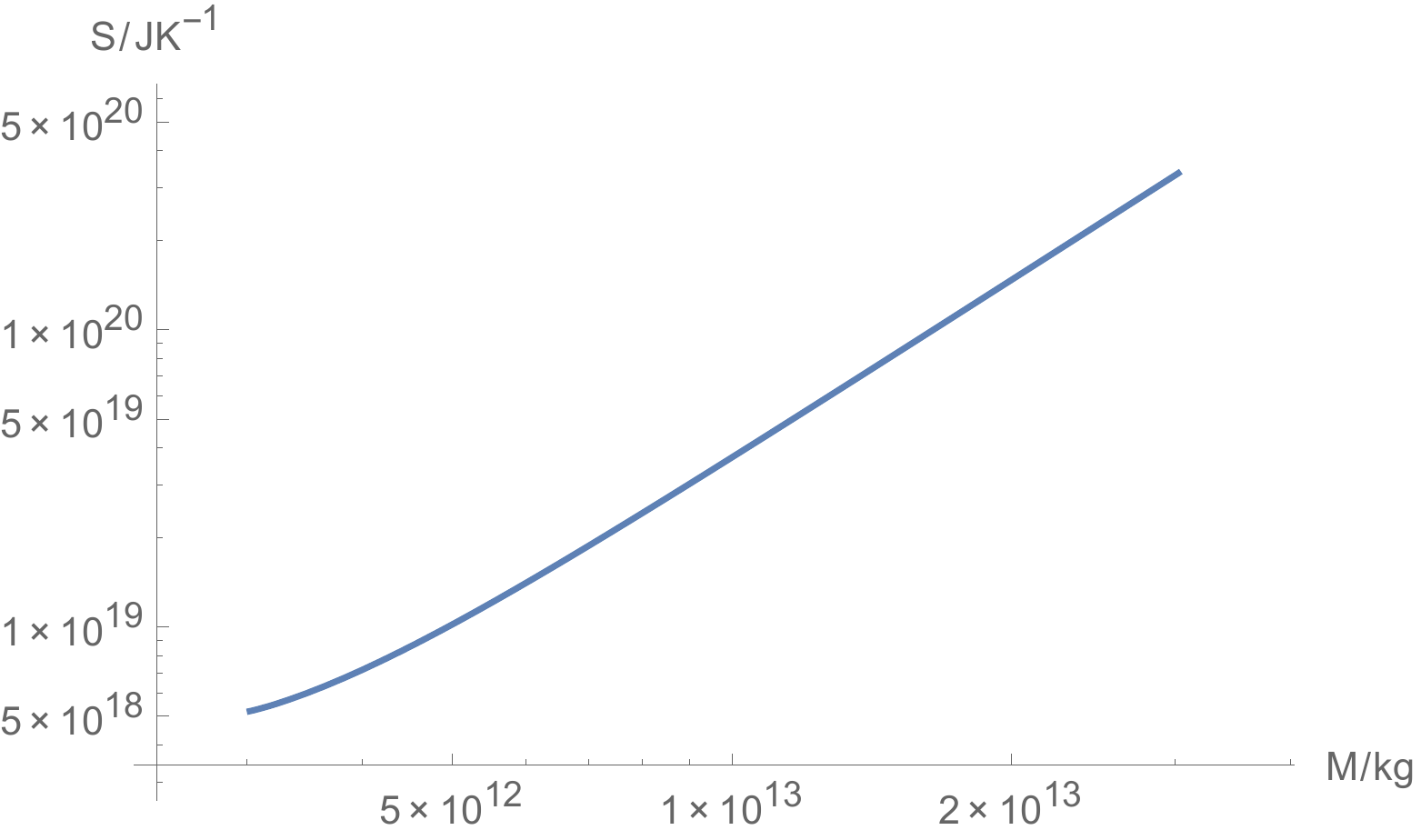}}
\caption{$T-M$ and $S-M$ relations for $R=R_{W}$.}
\label{fig2}
\end{figure}

The final value of mass can be determined from
\begin{eqnarray}
	1-\frac{R\hbar c^{3}}{4G^{2}M^{3}}&=&0,
\end{eqnarray}
leading to
\begin{eqnarray}
	M_{0}&=&\left(\frac{R\hbar c^{3}}{4G^{2}}\right)^{1/3}=\left(\frac{R\sqrt{\Lambda}}{4}\right)^{1/3}M'_{T}=\frac{1}{2^{2/3}}\left(\frac{R}{R_{W}}\right)^{1/3}M'_{T}=2^{1/3}M_{c}.  \label{Mrem}
\end{eqnarray}
It is no surprise that the black hole stops evaporate around the critical mass $M_{c}$ where the uncertainty in the position of emitted photons is minimum, i.e. $\Delta x_{min}\simeq (R_{p}^{2}R)^{1/3}$.  Interestingly, the temperature of the black hole remnant is still positive,
\begin{equation}
T_{0}=\frac{GM_{0}^{2}}{\pi Rk_{B}}=\frac{c^{2}}{\pi k_{B}}\left( \frac{\hbar^{2}}{16GR} \right)^{1/3}.
\end{equation}  
The value of the remnant temperature $T_{0}$ ranges from the Planck temperature $T_{p}= M_{p}c^{2}/k_{B}$~(for $R=R_{p}$) to the temperature $T_{T}$~(for $R=R_{W}$) associated with the new scale $M_{T}$,
\begin{equation} 
T_{T}=\frac{M_{T}c^{2}}{4^{2/3}\pi k_{B}}\simeq 10^{11}~{\rm K}.
\end{equation}
However, since the radiation flux is zero, this positive remnant temperature will be the last temperature of the black hole before the Hawking radiation stops.  A black hole that has radiated for a longer period of time will have higher uncertainty in $\Delta x$.  For black hole with the age comparable to $R_{W}/c$, it would stop radiating at mass around $M'_{T}$ when the temperature is $T_{T}$.

The additional uncertainty $2R\Delta p/Mc$ of the MLUR in (\ref{MLUReqn}) when applied to Hawking radiation can also be interpreted in the following way.  When a black hole emits a particle, we will never know the exact kicked-back momentum of the black hole until the emitted particle is detected, i.e. its wave function has collapsed.  Thus, the backreaction from momentum conservation will cause additional uncertainty in its position in the order of $t(\Delta v)$ where $t\sim 2R/c$ for a distance $R$ between black hole and the observer.  This is precisely the additional term in (\ref{MLUReqn}).  The longer a black hole has  been radiating, the larger the uncertainty generated by the backreaction on the position of the black hole.  

We can calculate the thermodynamic entropy of the black hole subject to the MLUR from the temperature function in Eqn.~(\ref{TMeqn}),
\begin{eqnarray}
S&=&c^{2}\int T^{-1} dM=\frac{3 B \, _2F_1\left(\frac{1}{3},\frac{1}{2};\frac{4}{3};\frac{B}{M^3}\right)+2 M \left(\sqrt{M^4-B M}+M^2\right)}{4 A B M/c^{2}},
\end{eqnarray}
where
\begin{equation}
A = \frac{G}{\pi k_{B}R}, \quad B=\frac{R\hbar c^{3}}{4G^{2}}, \quad AB=\frac{\hbar c^{3}}{4\pi G k_{B}}.
\end{equation}

The final state remnant with $M_{0}=B^{1/3}$ has entropy
\begin{equation}
S_{0}=\frac{c^{2}}{4 A \sqrt[3]{B}}\left( 2+\frac{3 \sqrt{\pi } ~\Gamma \left(\frac{4}{3}\right)}{\Gamma \left(\frac{5}{6}\right)}\right)=\frac{\pi k_{B}}{4^{2/3}}\left( 2+\frac{3 \sqrt{\pi } ~\Gamma \left(\frac{4}{3}\right)}{\Gamma \left(\frac{5}{6}\right)}\right)\frac{(\Delta x_{min})^{2}}{R_{p}^{2}},
\end{equation}
where $\Delta x_{min}=(R_{p}^{2}R)^{1/3}$.  Remarkably, the remnant entropy obeys holographic relation, i.e. it is proportional to the surface area of the remnant black hole $\sim (\Delta x_{min})^{2}$ in the unit of Planck area $R_{p}^{2}$.  Interestingly, even the uncertainty relation is modified to MLUR, holography is still present~(see also Ref.~\cite{Burikham:2015sro}).

\begin{figure}
	\centering
	\includegraphics{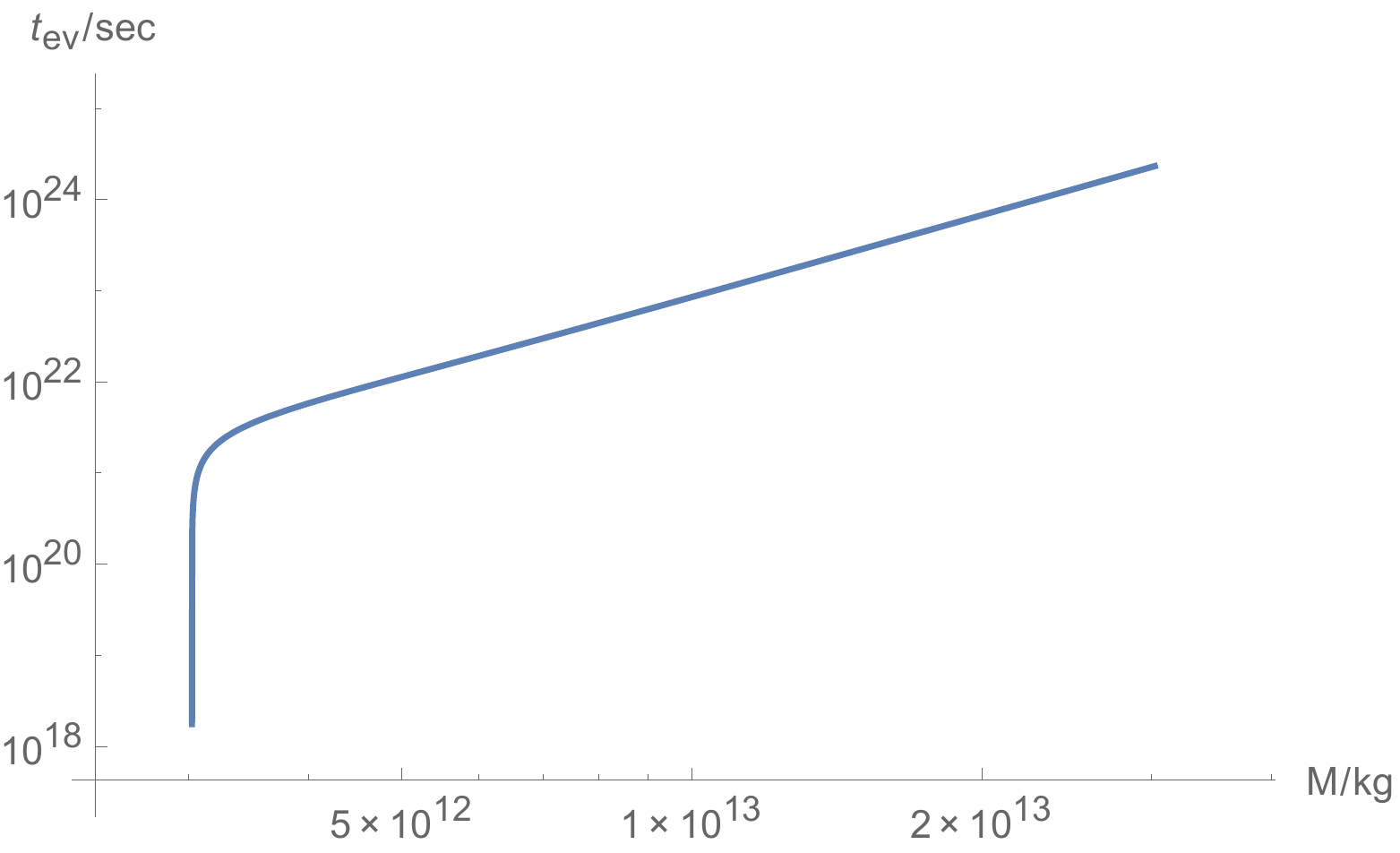}
	\caption{Evaporation lifetime of black hole for $R=R_{W}$.} \label{fig3}
\end{figure}

By assuming the Stefan-Boltzmann law of radiation, 
\begin{equation}
A_{h}\epsilon \sigma T^{4}=-c^{2}\frac{dM}{dt},
\end{equation}
the evaporation time of black hole until it reaches the final state remnant can be calculated,
\begin{eqnarray}
t_{ev}&=&\int_{B^{1/3}}^{M} \frac{c^{6}}{16\pi \epsilon \sigma T^{4}(GM)^{2}} dM,  \notag \\
         &=&\frac{16 \pi ^3 G^2 k_{B}^4}{3 c^6 \hbar^4 \sigma  \epsilon ~M^{3}}\Big[ -B^2+B M^3 \left( 8\sqrt{1-\frac{B}{M^3}}-7\right)+8 M^6 \left(\sqrt{1-\frac{B}{M^3}}+1\right)  \notag \\
         &&+8 B M^3 \left(2 \ln \left(\sqrt{M^3-B}+M^{3/2}\right)-3 \ln (M)\right)\Big].
\end{eqnarray}
If we set $R=R_{W},\epsilon =1$ and calculate the evaporation time of black hole with masses larger than the remnant mass $B^{1/3}$, it turns out that any black hole will live longer than the present age of the universe, 13.8 billion years$\simeq 4.35\times 10^{17}$ secs as shown in Fig.~\ref{fig3}.  The temperature and entropy to mass relations are presented in Fig.~\ref{fig2}.

\subsection{Minimum remnant mass}

Since the apparatus size cannot be smaller than the black hole, if we set the parameter $R=R_{min}=2GM/c^{2}=R_{S}$, the Schwarzschild radius of the {\it original} black hole at the beginning of the evaporation process.  Then the remnant mass in Eqn.~(\ref{Mrem}) becomes
\begin{equation}
M_{0}\simeq (M_{p}^{2}M)^{1/3},
\end{equation}
the weighed geometric mean of the original mass $M$ with the Planck mass.  And the remnant entropy is  
\begin{equation}
S_{0}=\pi k_{B}\left( \frac{M}{2M_{p}}\right)^{2/3}\left( 2+\frac{3 \sqrt{\pi } ~\Gamma \left(\frac{4}{3}\right)}{\Gamma \left(\frac{5}{6}\right)}\right)\simeq 6.207~\pi k_{B}\left( \frac{M}{2M_{p}}\right)^{2/3}.
\end{equation}
Therefore, a black hole starting with larger mass than $M_{p}$ will leave a remnant with a mass larger than $M_{p}$ as well.  Right before the black hole stops radiating, it has the final temperature equal to
\begin{equation}
T_{0}=\frac{M_{p}c^{2}}{2^{5/3}\pi k_{B}}\left( \frac{M_{p}}{M}\right)^{1/3}.
\end{equation}  

\section{A Generalization to $D$-dimensions and Holography}

In this section, we repeat the calculation of temperature and entropy of radiating black hole in noncompact $d$ dimensional spacetime.  It will become clear that the thermal entropy of the remnant black hole calculated from the MLUR is holographic in nature in arbitrary dimensions.  Starting with the MLUR
\begin{eqnarray}
	\Delta p&=&\frac{Mc\Delta x}{4R}\left( 1\pm \sqrt{1-\frac{4R\hbar}{Mc(\Delta x)^{2}}}\right),
\end{eqnarray}
we set $\Delta x = 2\chi M^{1/(d-3)}$, the Schwarzschild radius of a black hole in $d$ dimensions.  The gravitational constant $\chi = (\kappa c^{2})^{1/(d-3)}$ is dimension dependent with $\kappa = 8\pi G_{d}/c^{4}$.  By identifying the Hawking temperature via $\Delta p =\pi k_{B} T/c$, we obtain
\begin{eqnarray}
T &=& A_{d} M^{(d-2)/(d-3)}\left( 1-\sqrt{1-\frac{B_{d}}{M^{(d-1)/(d-3)}}} \right),
\end{eqnarray}
for
\begin{eqnarray}
A_{d}=\frac{c^{2}\chi}{2\pi k_{B}R},\quad B_{d}=\frac{\hbar R}{c\chi^{2}}.
\end{eqnarray}
The black hole radiation should stop when
\begin{equation}
M = B_{d}^{(d-3)/(d-1)},  \label{Mdeqn}
\end{equation}
where the size of the final black hole is roughly the minimum position uncertainty determined by the MLUR,
\begin{equation}
\Delta x = \Delta x_{min} \sim (\zeta^{2}\chi^{d-3})^{1/(d-1)},
\end{equation}
and $\zeta = \sqrt{R\hbar/c}$~\cite{Burikham:2015sro}.  The thermal entropy of the black hole can be calculated straightforwardly
\begin{eqnarray}
S&=&c^{2}\int T^{-1} dM, \\
 &=& \frac{(d-3) M^{-\frac{1}{d-3}} \left[ B_{d} (d-1) \, _2F_1\left(\frac{1}{2},\frac{1}{d-1};\frac{d}{d-1};B_{d} M^{-1-\frac{2}{d-3}}\right)+2 M^{\frac{d-1}{d-3}} \left(\sqrt{1-B_{d} M^{-\frac{2}{d-3}-1}}+1\right)\right]}{2 A_{d} B_{d} (d-2)}. \nonumber
\end{eqnarray}
When the radiation stops at the mass given by Eqn.~(\ref{Mdeqn}), the remnant entropy becomes
\begin{eqnarray}
S_{0}&=&\frac{c^{2}}{A_{d}B_{d}^{1/(d-1)}}\left( \frac{d-3}{d-2} \right)\left( 1+ \frac{\sqrt{\pi}}{2}\frac{\Gamma(1/(d-1))}{\Gamma((d+1)/2(d-1))} \right),  \\
&\sim& k_{B}\frac{(\Delta x_{min})^{d-2}}{R_{p,d}^{d-2}},
\end{eqnarray}
where $R_{p,d}=(\hbar \kappa c)^{1/(d-2)}$ is the $d$-dimensional Planck length.
The remnant black hole entropy is proportional to the horizon area, $(\Delta x_{min})^{d-2}$, in the unit of $d$-dimensional Planck area, $R_{p,d}^{d-2}$.  Remarkably, holography in thermal entropy of the remnant black hole persists in noncompact $d$ dimensions even when the uncertainty relation is modified to the MLUR~(conventional uncertainty relation with black hole physics also leads to holography as reflected in the area law of thermal entropy $S\sim M^{2}\sim R_{S}^{2}$ in 4 dimensions).  This is a strong evidence that holography is an inevitable phenomenon of the MLUR as discussed in Ref.~\cite{Burikham:2015sro}.

\section{Discussions and Conclusions}

We have proposed a new mass scale $M_{T}=(\hbar^{2}\sqrt{\Lambda}/G)^{1/3}$ constructed from 3 fundamental constants, $G, \hbar, \Lambda$.  In association to $M_{T}$, another mass scale $M'_{T}=c(\hbar/G^{2}\sqrt{\Lambda})^{1/3}$ can be defined which altogether divide the range of mass between the two Wesson masses into 3 equal parts in logarithmic scale.  Both new mass scales have physical interpretation in terms of the MLUR.  $M_{T}$ is the corresponding Compton mass of the minimum position uncertainty $\Delta x_{min}=(R_{p}^{2}R_{W})^{1/3}=R_{T}$ when observer is at cosmic horizon distance from the measured object whilst the critical mass giving $\Delta x_{min}=R_{T}$ is $M'_{T}$.    

When the MLUR is adopted to black hole radiation, it predicts remnant as the final state of the black hole evaporation process.  For a black hole with age at least around $R_{W}/c$, it will radiate until its temperature reaches as high as $M_{T}c^{2}/\pi k_{B}$ where its mass is of order of $M'_{T}$~(if it starts out at mass higher than $M'_{T}$).  For a general black hole, the {\it minimum remnant mass} is the weighed geometric mean $(M_{p}^{2}M)^{1/3}$ between the Planck mass and the original black hole mass $M$.  This occurs when $R$ is set to the minimal value $R_{min}=R_{S}=2GM/c^{2}$, the Schwarzschild radius of the original black hole.  The lowest possible remnant mass is the Planck mass only when the starting black hole is also of Planck mass according to the MLUR.  This is the crucial difference of the MLUR from the string-inspired GUP considered in Ref.~\cite{Adler:2001vs}.  According to MLUR, during the Hawking radiation process, the backreaction would give a black hole additional position uncertainty no less than the Schwarzschild radius of the original black hole.  The longer a black hole has been radiating, the larger the uncertainty in its position and size.  This kind of ``quantum random walk'' or quantum jitters results in the minimum size of the final remnant black hole $\sim (R_{p}^{2}R)^{1/3}$ which is larger than $R_{p}$~(for $R>R_{p}$).    

MLUR is a result of the intrinsic quantum uncertainty and gravitational uncertainty due to the horizon combined with the uncertainty originated in the measurement process by an observer at finite distance from the particle, or by an observer in an apparatus of a finite size.  The probe particle or detector that the observer uses to detect is in an entangled state with the measured particle until the measurement is finished.  Once the wave function is collapsed, the observer obtains the information of the measured particle and the probe.  However, during the measurement process while the particle is entangled with the probe, the particle uncertainty increases with time.  As a result, a minimum uncertainty in position of the particle becomes nonzero.  It can be identified with the Compton radius $R_{T}$~(roughly a femtometer) of the mass $M_{T}$ when the particle is entangled with the entire observable universe.  MLUR thus contains some aspect of quantum entanglement whence applied to black hole radiation~(i.e. entanglement between black hole and emitted radiation particle) leads to prediction of remnant final state of black hole.  It also provides another physical interpretation of the mass scales $M_{T}$ and $M'_{T}$ in terms of universe-age remnant temperature and mass.

Finally, in the scenario of black hole radiation where the MLUR is adopted, the black hole remnant has thermal entropy with holographic nature, i.e. proportional to its boundary area in the unit of Planck area.  This is proven in arbitrary noncompact dimensions.  Another unique aspect of the black hole evaporation according to Eqn.~(\ref{TMeqn}) is if we interpret the age of a black hole $t_{age}=R/c$ as the time it has been evaporating, the black hole temperature will depend on both the mass and its age $t_{age}$.  
\begin{equation}
	T=\frac{GM^{2}}{\pi k_{B}c t_{age}}\left(1-\sqrt{1-\frac{\hbar c^{4}t_{age}}{4G^{2}M^{3}}}\right).  \label{Ageeqn}
\end{equation}
For small $t_{age}$, the dependence on $t_{age}$ is very weak but becomes increasingly distinctive for large $t_{age}$.  The larger the age, the higher the temperature.  The black hole will finally stop evaporating at age $t_{age}=4G^{2}M^{3}/\hbar c^{4}$ and become a remnant.  

Remnants can be served as a dark matter candidate or a fraction of the necessary amount required by the observations.  Since remnants are weakly interacting and massive, they are categorized as a Cold-Dark-Matter~(CDM) candidate.  From Fig.~\ref{fig3}, black holes with masses above $10^{12}$ kg would remain until today.  The black hole more massive than $4.5\times 10^{22}$ kg will have lower temperature than the Cosmic-Microwave-Background temperature 2.73 K and they will keep growing.  Black holes with mass below $10^{12}$ kg would remain in the form of remnants.  In Ref.~\cite{Carr:2016hva}, constraints from Galactic and extra-galactic gamma-ray background on the population of the black holes with masses around $10^{12}$ kg are estimated.  To avoid the over population of the remnants, an era of inflation is required~\cite{Chen:2002tu}.          

\acknowledgments

We would like to thank Matt Lake for multiple useful discussions.  P.B. and T.W. are supported in part by the Thailand Research Fund~(TRF), Commission on Higher Education~(CHE) and Chulalongkorn University under grant RSA5780002.

\appendix  

\section{Relationship between Wesson masses and Hawking temperature}     \label{app}

In section \ref{SectTH}, we have shown that the mass scale $M_{T}, M'_{T}$ emerge from the Hawking radiation when the uncertainty relation is modified to the MLUR and the apparatus size to is set to $R=R_{W}$, the universe size.  Naturally, we would wonder if conventional black hole thermodynamics in the finite size universe $R_{W}$ would yield the similar results.  

In this section, we will show that the Hawking temperature of a black hole in de Sitter~(dS) and Anti de Sitter~(AdS) space with cosmic radius $R_{W}$ can be expressed in infinite series involving only the 2 Wesson masses $M_{W},M'_{W}$.  Therefore, the main results that $M_{T}, M'_{T}$ emerge from the Hawking radiation are very unique characteristic of the MLUR~(with $R=R_{W}$) which cannot be reproduced by conventional black hole thermodynamics and uncertainty relation.  

In the Schwarzschild-(Anti) de Sitter space, the metric with $\sqrt{1/\Lambda}=R_{W}$ substituted is,
\begin{equation}
	ds^{2}=-\left(1-\frac{2GM}{c^{2}r}\pm\frac{\Lambda}{3}r^{2}\right)dt^{2}+\left(1-\frac{2GM}{c^{2}r}\pm\frac{\Lambda}{3}r^{2}\right)^{-1}dr^{2}+r^{2}d\Omega^{2},\nonumber
\end{equation}
where the plus~(minus) sign is for the AdS~(dS) case.
We can find the horizon radius by setting $\left(1-\frac{2GM}{c^{2}r}\pm\frac{\Lambda}{3}r^{2}\right)$ equal to zero giving~\cite{Rahman:2012id,Cruz:2004ts}
\begin{eqnarray}
	r_{dS}&=&\frac{2}{\sqrt{\Lambda}}\cos\left[\frac{\pi}{3}+\frac{1}{3}\cos^{-1}\left(3\frac{GM\sqrt{\Lambda}}{c^{2}}\right)\right], \label{a1} \\
	r_{AdS}&=&\frac{2}{\sqrt{\Lambda}}\sinh\left[\frac{1}{3}\sinh^{-1}\left(3\frac{GM\sqrt{\Lambda}}{c^{2}}\right)\right],   \label{a2}
\end{eqnarray}
where $r_{dS}$ and $r_{AdS}$ denote horizon radius in the Schwarzschild-de Sitter and Schwarzschild-Anti-de Sitter space respectively.

Thus, the black hole temperature is,
\begin{eqnarray}
	T=\frac{\hbar c}{k_{B}}\frac{g'_{00}(r_{h})}{4\pi}=\frac{\hbar c}{4\pi k_{B}r_{h}}\left( 1\pm \Lambda r_{h}^{2}\right),   \label{a3}
\end{eqnarray}	
where $r_{h}=r_{AdS}~(r_{dS})$ for $+~(-)$ sign.  Eqn.~(\ref{a3}), together with (\ref{a1}) and (\ref{a2}), can be expanded to yield
\begin{eqnarray}
	T_{dS}&=&\frac{M_{W}c^{2}}{8\pi k_{B}}\left(\frac{M'_{W}}{M}-16\frac{M}{M'_{W}}-80\left(\frac{M}{M'_{W}}\right)^3-...\right),\nonumber \\
	T_{AdS}&=&\frac{M_{W}c^{2}}{8\pi k_{B}}\left(\frac{M'_{W}}{M}+16\frac{M}{M'_{W}}-80\left(\frac{M}{M'_{W}}\right)^3+...\right).\nonumber
\end{eqnarray}
Although this result does not contain any relation with our main mass scales, $M_{T}$ and $M'_{T}$, it is quite interesting to see the two Wesson masses showing up in the Hawking temperature formulae.

\end{document}